\documentclass[fleqn,usenatbib,useAMS]{mnras}

\usepackage[T1]{fontenc}
\usepackage{graphicx}
\usepackage{amsmath}
\usepackage{epsfig,amsmath,natbib,mathtools}
\usepackage{amssymb}
\usepackage{aas_macros}
\usepackage{tikz,hyperref}
\usepackage{xcolor}
\colorlet{mylinkcolor}{violet}
\colorlet{mycitecolor}{blue}
\colorlet{myurlcolor}{magenta}
\usepackage{soul}
\hypersetup{
	linkcolor  = mylinkcolor,
	citecolor  = mycitecolor,
	urlcolor   = myurlcolor,
	colorlinks = true,
}
\usepackage{orcidlink}

\DeclareRobustCommand{\okina}{%
  \raisebox{\dimexpr\fontcharht\font`A-\height}{%
    \scalebox{0.8}{`}%
  }%
}

\title[Indicator Function Information]{Indicator Functions: Distilling the Information from Gaussian Random Fields}

\author[A. Repp et al.]{
Andrew Repp\orcidlink{0000-0002-0116-1039}$^{1}$
Ravi K. Sheth\orcidlink{0000-0002-2330-0917}$^{2}$
Istv\'an Szapudi\orcidlink{0000-0003-2274-0301}$^{3}$
and Yan-Chuan Cai\orcidlink{0000-0002-9855-2342}$^{4}$\thanks{cai@roe.ac.uk}
\\
$^{1}$Christian Liberty Academy, Kea{\okina}au, HI 96749, USA\\
$^{2}$Department of Physics and Astronomy, University of Pennsylvania, Philadelphia, PA 19104, USA\\
$^{3}$Institute for Astronomy, University of Hawaii, 2680 Woodlawn Drive, Honolulu, HI 96822, USA\\
$^{4}$Institute for Astronomy, University of Edinburgh, Royal Observatory, Blackford Hill, Edinburgh, EH9 3HJ, UK
}

\date{\today; submitted to MNRAS}
\begin{document}
\label{firstpage}
\pagerange{\pageref{firstpage}--\pageref{lastpage}}
\maketitle

\begin{abstract}
A random Gaussian density field contains a fixed amount of Fisher information on the amplitude of its power spectrum. For a given smoothing scale, however, that information is not evenly distributed throughout the smoothed field. We investigate which parts of the field contain the most information by smoothing and splitting the field into different levels of density (using the formalism of indicator functions), deriving analytic expressions for the information content of each density bin in the joint-probability distribution (given a distance separation). When we choose one particular distance regime (i.e., cells separated by $60$--$80h^{-1}$ Mpc), we find that the information in that range peaks at moderately rare densities (where the number of smoothed survey cells is roughly of order of magnitude 100). Counter-intuitively, we find that, for a finite survey volume (again at a particular distance range), indicator function analysis can outperform conventional two-point statistics while using only a fraction of the total survey cells, and we explain why. In light of recent developments in marked statistics (such as the indicator power spectrum and density-split clustering), this result elucidates how to optimize sampling for effective extraction of cosmological information. 
\end{abstract}


\section{Introduction}
The $\Lambda$CDM framework neatly explains the bulk of available cosmological data (e.g., \citealp{Planck2018}), and a major goal of current and future surveys is to measure these parameters more precisely.
Such surveys include CMB-mapping efforts like CMB-S4 \citep{Abazajian2019}; line intensity surveys like CHIME (Canadian Hydrogen Intensity Mapping Experiment, \citealp{Newburgh2014}) and COMAP (CO Mapping Array Project, \citealp{Kovetz2017}); and galaxy surveys like DESI (Dark Energy Spectroscopic Instrument, \citealp{DESI2016}), \emph{Euclid} \citep{Laureijs2011} and \emph{Roman} \citep{Akeson2019}.

Fisher information provides a useful metric for the constraining power of these surveys. Unfortunately, more data does not necessarily translate into more information. In particular, the standard analysis tools (i.e., power spectrum and correlation function) exhibit an information plateau at higher wavenumbers \citep{Neyrinck2007, Lee2008, Wolk2015}.

Multiple strategies exist for circumventing this plateau. One is to apply Gaussianizing transformations (e.g., the log transform,   \citealp{Neyrinck2009}, given the approximate lognormality of the matter field). Another strategy is to
condition power spectrum analysis on the local-density environment (measured through, e.g., counts-in-cells). This combination (power spectrum plus counts-in-cells) represents a variety of mark-correlation analysis \citep{Sheth1998,Beisbart2000,Szapudi2000,Skibba2006}, with densities serving as marks \citep{Martino2009,White2009, Chiang2015,White2016, Gruen2016,Gruen2018, Massara2021, Paillas2021, Paillas2023}; see in particular the pioneering work of \citet{AbbasSheth2005,AbbasSheth2007}, as well as \citet{Paranjape2018,Shi2018,Alam2019}. In addition, since each density bin defines its own set of regions in a survey volume, the simultaneous use of multiple density bins is a species of multitracer analysis \citep{McDonald2009,Bernstein2011, Hamaus2011,Barreira2023,Nikakhtar2024}.

\citet{ReppSzapudi2022} introduced indicator functions as a unifying framework for such density-dependent analyses. This framework enables us to understand where the survey's information comes from, facilitating the design of more efficient information-extraction methods. In this work, we apply indicator functions to Gaussian random fields and demonstrate analytically that a large amount of the information typically resides in a small number of density extrema. Given the importance of optimal weighting (e.g., \citealp{Cai2011}), it is significant that we can focus our attention on the most information-rich (as opposed to noise-dominated) survey cells, thus distilling out the bulk of the field's information. 
Such information-locative knowledge should increase the efficiency of survey data analysis, as well as assisting formulation of measures which are more robust to various kinds of systematics.

We focus this work on the amplitude of the power spectrum; thus, in this paper, ``information'' denotes specifically information on this amplitude; however, the analysis should apply to any amplitude-like parameter. It might not apply to shape parameters, since we assume that we already know the shape of the power spectrum.

We organize the paper as follows: in Section~\ref{sec:totinfo} we derive an expression for the total amount of information in an $n$-point Gaussian field. Section~\ref{sec:indfunc} defines indicator functions and their correlations. Section~\ref{sec:indinfo} then provides analytic expressions for the information in indicator correlations for a Gaussian field. In Section~\ref{sec:infotest}  we test the accuracy of our expressions, with discussion and conclusion occupying the final two sections.

\section{Information in a Gaussian Distribution}
\label{sec:totinfo}
We start by establishing the total amount of information (on $\sigma^2$, the fluctuation amplitude) contained in a Gaussian field.

Consider first a one-point probability distribution function $\mathcal{P}(\delta)$ with zero mean and variance $\sigma^2$, $\delta$ being the (real-space) overdensity in such a field. The Fisher information on $\ln\sigma^2$ is 
\begin{equation}
  I_1 \equiv \int d\delta\,\mathcal{P}(\delta)\,\left(\frac{d\ln \mathcal{P}(\delta)}{d\ln\sigma^2}\right)^2
\end{equation}
For a Gaussian distribution, 
\begin{equation}
  \mathcal{P}(\delta) = \frac{\exp(-\delta^2/2\sigma^2)}{\sqrt{2\pi\sigma^2}}.
\end{equation}
Thus, if we define $\nu\equiv\delta/\sigma$, the information 
 $I_1 = \langle(\nu^2 - 1)^2/2^2\rangle = 1/2.$

Now consider the two-point distribution
\begin{equation}
    \mathcal{P}_2(\delta_1,\delta_2) = \mathcal{P}(\delta_1)\, \mathcal{P}(\delta_2|\delta_1),
\end{equation}
and suppose that both one point distributions $\mathcal{P}(\delta_1)$ and $\mathcal{P}(\delta_2)$ have mean zero and variance $\sigma^2$.  The Fisher information on $\ln\sigma^2$ is now
\begin{align}
    I_2 &\equiv \int d\delta_1\, d\delta_2\,\mathcal{P}(\delta_1,\delta_2)\, 
      \left(\frac{d\ln \mathcal{P}_2}{d\ln\sigma^2}\right)^2 \nonumber\\
    &= \int d\delta_1\,\mathcal{P}(\delta_1) \int d\delta_2\,\mathcal{P}(\delta_2|\delta_1)\times\nonumber\\
    &\hspace{3cm}\nonumber\left(\frac{d\ln \mathcal{P}
    (\delta_1)}{d\ln\sigma^2} + 
             \frac{d\ln \mathcal{P}(\delta_2|\delta_1)}{d\ln\sigma^2}\right)^2 \nonumber\\
    &= I_1 + \int d\delta_1\,\mathcal{P}(\delta_1)\,I_{2|1} .
\end{align}
For a bivariate Gaussian, $\mathcal{P}(\delta_2|\delta_1)$ is also Gaussian, with mean 
 $\mu_{2|1} \equiv\delta_1\,\langle\delta_2\delta_1\rangle/\sigma^2$ and variance 
 $\sigma^2\, (1 - \gamma^2)$, where $\gamma = \langle\delta_2\delta_1\rangle/\sigma^2$.
The amplitude of the power spectrum cancels out in the ratios defining $\mu_{2|1}$ and $\gamma$.  So, if we let $\nu_{2|1} = (\nu_2 - \gamma\nu_1)/\sqrt{1-\gamma^2}$, then 
 $I_{2|1} = \langle(\nu_{2|1}^2 - 1)^2/4\rangle = 1/2$, making $I_2 = 2I_1.$

By repeating this logic, it is straightforward to see that the information on $\ln\sigma^2$ contained in an $n$-point Gaussian distribution generalizes to $I_n = nI_1 = n/2$.  

We can extend this result to lognormal distributions. If we define $A = \ln(1+\delta)$ and assume $A$ is Gaussian, then 
\begin{equation}
 \mathcal{P}(A) = \frac{{\rm e}^{-(A + \sigma_A^2/2)^2/2\sigma_A^2}}{\sqrt{2\pi \sigma^2_A}},
\end{equation}
with $\langle A \rangle = -\sigma^2_A/2$. Since
 $1 + \sigma_{\delta}^2 = \exp(\sigma^2_A)$,
we must decide if we are interested in the amplitude of $P_A(k)$ or of $P_\delta(k)$.  The form above suggests that it is more convenient to work with $\sigma_A^2$, since then the Fisher information is 
\begin{equation}
    I_1 = \langle [\nu_A^2/2 - 1/2-\sigma_A\nu_A/2]^2\rangle
        = (1 + \sigma^2_A/2) /2
    \label{eq:info1pixel}
\end{equation}
where $\nu_A\equiv [A - \langle A \rangle]/\sigma_A = [A + \sigma_A^2/2]/\sigma_A$.  
We can then calculate the Fisher information on $\ln \sigma_\delta^2$ from that on $\ln \sigma_A^2$, given that they are related by the square of the Jacobian
 $d\ln\sigma^2_\delta/d\ln\sigma^2_A = (\sigma^2_A/\sigma^2_\delta)(1+\sigma^2_\delta)$.  
Since a lognormal is just a monotonic mapping of a Gaussian, one might have thought that the information in the lognormal is the same as that in the Gaussian.  The apparent paradox is resolved by noting that the Gaussian variable which transforms to the lognormal (i.e., $A$) has non-zero mean, and this mean depends on $\sigma^2_A$.  Thus, the lognormal provides information about $\sigma^2_A$ both through the mean of $A$ and through its variance.  Indeed, the expression above tends to the Gaussian information as $\sigma_A^2$ approaches $0$,  consistent with the fact that the lognormal PDF approaches the (zero-mean) Gaussian in this limit. However, $I_2 = 2I_1$ just as for the Gaussian.  

\section{Indicator Functions}
\label{sec:indfunc}
Density-split statistics (e.g., the density-split clustering analysis in BOSS CMASS, \citealp{Paillas2021,Paillas2023,CuestaLazaro2024,Morawetz2025}, and in DESI, \citealp{Gruen2016,Friedrich2018,Gruen2018}) is well-formulated in terms of indicator functions, which we now  define.

Following \citet{ReppSzapudi2022}, we say that any bin of values $B$ defines a corresponding \emph{indicator function} on a random field $X$, so that at every point $x \in X$, the indicator function is unity if and only if $x \in B$. Thus, $I_B:X\rightarrow \{0,1\}$, such that
\begin{equation}
\mathcal{I}_B(x) = \begin{cases} 
      1 & x \in B \\
      0 & \mbox{otherwise}.
   \end{cases}
\end{equation}
Note that this indicator function is identical to the characteristic function $\chi_i$ of \citet{Bernardeau2022}.

For $X$ we use the density contrast field $\delta(\mathbf{r}) = \rho(\mathbf{r})/\overline{\rho} - 1$, with $\rho$ being the density at $\mathbf{r}$ (and $\overline{\rho}$ being the mean density). Then for any density bin $B$, the indicator function $\mathcal{I}_B$ has a value of 1 wherever $\delta \in B$, but vanishes at all other points. Clearly the expected value of the indicator function equals the probability of the bin $\overline{\mathcal{I}}_B = \mathcal{P}(B)$.

We can further normalize the indicator function ($\delta_{\mathcal{I}_B} = \mathcal{I}/\overline{\mathcal{I}}_B - 1$) and define indicator function correlations\footnote{Closely related to the sliced correlations of \citet{Neyrinck2018}.} just as we do with the full random field: given a distance $r$,
\begin{align}
\xi_{I_B}(r) & \equiv \frac{\left\langle \delta_{\mathcal{I}_B}(\mathbf{x}_1)\delta_{\mathcal{I}_B}(\mathbf{x}_2) \right\rangle_{|\mathbf{x_2}-\mathbf{x_1}|=r}}{\mathcal{P}(B)^2} - 1 \\
& = \frac{\mathcal{P}_{11}(B)}{\mathcal{P}(B)^2} - 1,
\end{align}
where $\mathcal{P}_{11}(B)$ is the probability that two points separated by a distance of $r$ both fall into density bin $B$.

In the case of a Gaussian field with weak correlations (so that $\gamma \equiv \xi(r)/\sigma^2 \ll 1$), \citet{ReppSzapudi2022} derive the following relationship between the standard two-point correlation function $\xi(r)$ and an indicator correlation function $\xi_{I_B}(r)$:
\begin{equation}
    \xi_{I_B}(r) = \frac{\xi(r) \langle \nu \rangle_B^2}{\sigma^2},
    \label{eq:IFbias}
\end{equation}
where $\nu \equiv \delta/\sigma$, and the average is taken over all cells in bin $B$.

Note that these results are by no means \emph{de novo}. The above-referenced paper explores the close connection between Equation~\ref{eq:IFbias} and the results of \citet{Kaiser1984} (see also the seminal results of \citealt{BBKS}). Analogous and/or related work appears in \citet{Bernardeau1996}, who explores the bias properties of extreme-density regions, and in both \citet{Codis2016} and \citet{Uhlemann2017}, who consider bias expressions for statistics of density pdfs at separated locations---the latter demonstrating significant improvement in correlation function estimation. Other work (e.g., \citealp{Thiele2020}, \citealp{Bernardeau2022, Uhlemann2023}) expresses the covariance of one-point pdfs in terms of two-point statistics.

For the remainder of this paper, we suppress notation of the bin $B$ and write $\xi_I$, understanding that $I$ requires a specified bin.

\section{Information in Indicator Correlation Functions}
\label{sec:indinfo}
Now suppose we have a Gaussian random field which evolves in a linear manner. Let $A_z$ represent the amplitude of the power spectrum $P(k)$ with respect to some fiducial value (i.e., the primordial amplitude $A_s$ after being evolved to  reference time $z$). If $P_\textrm{fid}(k)$ is the fiducial power spectrum value, then
\begin{equation}
    A_z \equiv \frac{P(k)}{P_\textrm{fid}(k)} = \frac{\xi(r)}{\xi_\mathrm{fid}(r)},
\end{equation}
where $\xi(r)$ and $\xi_\mathrm{fid}(r)$ are the two-point correlation function and its fiducial value at distance $r$. Writing $\xi(r) = \gamma(r) \sigma^2$, we can say 
\begin{equation}
    A_z = \frac{\sigma^2}{\sigma_\mathrm{fid}^2}.
    \label{eq:Az}
\end{equation}
We want to calculate the information on $A_z$ contained in the indicator-function correlations $\xi_I(r)$, within a given interval of $r$-values.

The indicator correlation function itself is thus our observable. It has a theoretical value (the average value over an ensemble of surveys) for which we reserve the notation $\xi_I(r)$. The actual observed values of the indicator-function correlations will differ stochastically from survey to survey; we denote the observed value as $\hat{\xi}_I(r)$.

Determining the information in $\xi_I(r)$ requires calculation of the expected value of $((\partial/\partial A_z)\ln \hat{\xi}_I(r))^2$, which in turn requires calculation of the probability distribution of $\hat{\xi}_I(r)$. Thus, given a theoretical $\xi_I(r)$, we must determine the likelihood of observing a given $\hat{\xi}_I(r)$.

\subsection{Probability Distribution for $\hat{\xi}_I(r)$}
Suppose we have a survey consisting of $N_c$ cells (i.e., smoothed with a top-hat filter at discrete points); we choose a bin $B$ of densities to determine our indicator function. We also choose a bin $R$ of $r$-values for our correlation function.

Let $\hat{N}_1$ be the number of the cells on which the indicator function is unity (i.e., the number in density bin $B$). Let $N_p$ be the number of \emph{pairs} of cells (of any density) separated by a distance $r \in R$. Let $\hat{N}_{11}$ be the number of such pairs on which the indicator function for \emph{both} cells is unity.

Say the probability of a cell falling into density bin $B$ is $P_1$, and the probability of a pair of cells (separated by distance $r \in R$) both falling into $B$ is $P_{11}$. Then the expected value of the (normalized) indicator-function correlation is
\begin{equation}
    \xi_I(r)=\frac{P_{11}}{P_1^2}  - 1.
\end{equation}
However, the observed correlation is
\begin{equation}
    \hat{\xi}_I(r) = \frac{\hat{P}_{11}}{\hat{P}_1^2}  - 1,
    \label{eq:xihatdef}
\end{equation}
where $\hat{P}_1$ and $\hat{P}_{11}$ are the empirical probabilities $\hat{N}_1/N_c$ and $\hat{N}_{11}/N_p$, respectively. 
We want to determine the likelihood $\mathcal{P}(\hat{\xi}_I(r))$ given underlying probabilities $P_1$ out of $N_c$ cells, and given $P_{11}$ out of $N_p$ pairs, and we obtain this likelihood by differentiating the cumulative distribution function:

\begin{align}
    \mathcal{P}(\hat{\xi}_I) & = \frac{d}{d\hat{\xi}_I} \mathrm{Pr}(\hat{\xi}_I' \leq \hat{\xi}_I) \\
    & = \frac{d}{d\hat{\xi}_I} \int d\hat{P}_1 \int_{0}^{\hat{P}_1^2(1+\hat{\xi}_I)} d\hat{P}_{11}\, \mathrm{Pr}(\hat{P}_{11} | \hat{P}_1)\,\mathrm{Pr}(\hat{P}_1)\\
    & = \int d\hat{P}_1 \hat{P}_1^2 \, \mathrm{Pr}\!\left(\hat{P}_{11} = \hat{P}_1^2(1+\hat{\xi}_I) | \hat{P}_1\right) \mathrm{Pr}(\hat{P}_1).\label{eq:condprob}
\end{align}

To obtain an expression for this integral,\footnote{We cannot assume a binomial distribution since the $N_c$ cells are correlated.}
let us assume that the empirical probabilities $\hat{P}_1$ scatter about the expected value $P_1$ with variance $\sigma_1^2$, so that
\begin{equation}
\mathrm{Pr}\!\left(\hat{P}_1 = x\right) = G(x; P_1, \sigma_1^2);
\end{equation}
likewise, we assume a Gaussian distribution for the conditional probability in Equation~\ref{eq:condprob}:
\begin{equation}
\mathrm{Pr}\!\left(\hat{P}_{11} = y \,|\, \hat{P}_1 = x\right) = G(y; x^2(1+\xi_I), \sigma_{1|1}^2).
\label{eq:P1g1}
\end{equation}
In a moment we shall make an additional assumption about the relationship between $\sigma_1$ and $P_1$, but before doing so, we can express the distribution of $\mathcal{P}(\hat{\xi}_I(r))$ as follows:
\begin{multline}
    \mathcal{P}(\hat{\xi}_I) = \frac{1}{2\pi\sigma_{1|1}\sigma_1} \times\\
    \int dx\,x^2\, \exp\left(
    -\frac{x^4(\hat{\xi}_I-\xi_I)^2\sigma_1^2 + (x-P_1)^2\sigma_{1|1}^2}{2\sigma_{1|1}^2\sigma_1^2}
    \right).
\end{multline}

Performing this integration is nontrivial. For this reason we make the further assumption that $\sigma_1 \ll P_1$. Using the expression from  \citet{ReppSzapudi2021}, one can show that this assumption is equivalent to
\begin{equation}
    N_1 \gg \frac{1}{1-\overline{\xi}_I^{\neq}} \approx 1,
    \label{eq:N1big}
\end{equation}
where $\overline{\xi}_I^{\neq}$ is the volume-averaged indicator correlation function, with the average taken over all pairs of distinct survey cells. Thus, so as long as a decent number of survey cells occupy the density bin under consideration, this assumption should be valid.\footnote{Note that Equation~\ref{eq:N1big} requires $N_c \gg 1 \gg\overline{\xi}_I^{\neq}$, which is typically true.}

Under this assumption, the exponential is insignificant unless $x \approx P_1$. In this regime,
\begin{multline}
    \mathcal{P}(\hat{\xi}_I) \approx \frac{1}{2\pi\sigma_{1|1}\sigma_1} \int dx\,P_1^2\, \exp\left(
    -\frac{P_1^4(\hat{\xi}_I-\xi_I)^2}{2\sigma_{1|1}^2}\right)\\
    \times\exp\left(\frac{-(x-P_1)^2}{2\sigma_1^2}\right)
\end{multline}
so that
\begin{equation}
    \mathcal{P}(\hat{\xi}_I) \approx \frac{P_1^2}{\sigma_{1|1}\sqrt{2\pi}} \exp\left(
    -\frac{P_1^4(\hat{\xi}_I-\xi_I)^2}{2\sigma_{1|1}^2}\right).\label{eq:xiIdistrib}
\end{equation}
Thus, in this regime, the measured values of the indicator function correlations will be normally distributed with mean $\xi_I$ and variance $\sigma_{1|1}^2/P_1^4$.

We must now estimate $\sigma_{1|1}^2$, the variance of the (conditional) two-point probability measurement $\hat{P}_{11}$ in Equation~\ref{eq:P1g1}. We do so as follows.

In any realization of this field, the number of cells in density bin $B$ is $\hat{N}_1 = N_c \hat{P}_1$. Suppose that for any cell $x_0$ in the survey, there are $2k$ other cells a distance $r \in R$ from $x_0$.\footnote{This assumes periodic boundary conditions; otherwise it is a good approximation if $r$ is small compared to the survey size.} Then the total number of pairs of cells separated by $r \in R$ is $N_p = N_c k$ (dividing by 2 to avoid double-counting the pairs).

Now, we know there are $\hat{N}_1$ cells in density bin $B$. If we focus on the $2\hat{N}_1 k$ cells the requisite distance away from them, those cells will also be in $B$ with probability $P_1 (1 + \xi_I)$. Therefore, the total number of pairs of cells (separated by $r \in R$) for which the indicator function on both is unity is given by the binomial distribution:
\begin{equation}
    \hat{N}_{11} \sim \mathrm{Bin}(\hat{N}_1 k, P_1 (1 + \xi_I)).
\end{equation}

Thus we have the conditional variance of $\hat{P}_{11} = \hat{N}_{11}/N_p$:
\begin{align}
    \mathrm{Var}(\hat{P}_{11}|\hat{P}_1) & = \frac{1}{N_p^2}\mathrm{Var}(\hat{N}_{11}|\hat{N}_1)\\
    & = \frac{1}{N_p} \hat{P}_1 P_1 (1 + \xi_I) (1 - P_1 (1 + \xi_I)).
\end{align}

Therefore the expected value of this conditional variance (averaging across empirical probabilities $\hat{P}_1$) is
\begin{equation}
    \sigma_{1|1}^2 = \langle\mathrm{Var}(\hat{P}_{11}|\hat{P}_1)\rangle = \frac{P_1^2}{N_p}(1 + \xi_I)(1 - P_1(1 + \xi_I)).
    \label{eq:condvar}
\end{equation}

We note that Equation~\ref{eq:condvar} is still an approximation, since if two cells $x_1$ and $x_2$ are both a distance $r$ from cell $x_0$, the values of $x_1$ and $x_2$ are correlated, and strictly speaking the binomial distribution does not apply. In particular, there is a dependence on the value of $k$, such that as $k$ increases, the variance $\sigma_{1|1}^2$ will in fact exceed the value given in Equation~\ref{eq:condvar}.

However, if we assume this expression for $\sigma^2_{1|1}$, we can from Equation~\ref{eq:xiIdistrib} conclude that the variance of $\hat{\xi}_I$ is
\begin{equation}
    \sigma^2_{\hat{\xi}_I} = \frac{(1 + \xi_I)(1 - P_1(1 + \xi_I))}{P_1^2 N_p},\end{equation}
and if we work to first order in $\xi_I$, we can write
\begin{equation}
    \sigma^2_{\hat{\xi}_I} = \frac{1 - P_1 + \xi_I(1 - 2P_1)}{P_1^2 N_p},
    \label{eq:xiIvariance}
\end{equation}

For the expected value of the distribution, $\langle \hat{\xi}_I \rangle$ is simply the indicator correlation $\xi_I$. Using the expression in \citet{ReppSzapudi2022} (for Gaussian fields in the linear regime), we can write
\begin{equation}
    \langle \hat{\xi}_I \rangle = \frac{\xi\delta^2}{\sigma^4} = \frac{\gamma \delta^2}{A_z \sigma^2_\mathrm{fid}}.
    \label{eq:xiIexpression}
\end{equation}

We note again that $\gamma \equiv \xi/\sigma^2$, and that $\xi_I$ denotes the indicator correlation, while $\xi$ denotes the standard two-point correlation of the field $\delta$.

\subsection{Calculating Fisher Information}
Given this PDF for $\hat{\xi}_I$, we can now calculate the Fisher information on the amplitude $A_z$ contained in the observable $\hat{\xi}_I$, given a distance range $R$. We assume a Gaussian distribution for $\delta$, so that $P_1 = G(\delta; 0, \sigma^2)$.\footnote{In the linear regime for the cosmological matter distribution, $A_z \propto \sigma^2 = \sigma^2_\mathrm{lin}$; on smaller scales, one could adapt these results to the roughly-Gaussian log distribution of $A = \ln(1+\delta)$.}

Writing $\Delta$ for $\hat{\xi}_I - \xi_I$ and noting that $\sigma^2_{\hat{\xi}_I} = \langle \Delta^2 \rangle$, we have
\begin{equation}
    \mathcal{P}(\hat{\xi}_I) = \frac{1}{\sqrt{2\pi \langle \Delta^2 \rangle}} \exp\left(-\frac{\Delta^2}{2\langle \Delta^2 \rangle}\right).
\end{equation}
Then the information on $A_z$ is
\begin{align}
    \mathcal{I}_{\mathbf{A_z}} & = \left\langle \left(\frac{d}{dA_z} \ln \mathcal{P}(\hat{\xi}_I) \right)^2 \right\rangle \\
    \begin{split}
    & = \frac{(\langle \Delta^2 \rangle')^2}{4\langle \Delta^2 \rangle^2} + \langle\Delta^2\rangle \left( \frac{(\Delta')^2}{\langle \Delta^2 \rangle^2} -  \frac{(\langle \Delta^2 \rangle')^2}{2\langle \Delta^2 \rangle^3}\right) + \langle\Delta^4\rangle\frac{(\langle \Delta^2 \rangle')^2}{4\langle \Delta^2 \rangle^4}\\
    & \quad\quad + \left\langle
    \textrm{odd powers of $\Delta$}\right\rangle\\
    \end{split}\\
    & = \frac{1}{2}\left(\frac{\langle \Delta^2 \rangle'}{\langle \Delta^2 \rangle}\right)^2 + \frac{(\Delta')^2}{\langle \Delta^2 \rangle},
\end{align}
where primes denote derivatives with respect to $A_z$, and the odd moments vanish because of the normal distribution of $\hat{\xi}_I$. From Equation~\ref{eq:xiIexpression} we can write $\Delta'=\gamma\nu^2/A_z = $ (where again $\nu = \delta/\sigma$) and write the logarithmic derivative $L = (d/dA_z)(\ln \langle \Delta^2 \rangle)$ to obtain
\begin{equation}
    \mathcal{I}_{A_z} = \frac{L^2}{2} + \frac{\gamma^2\nu^4}{A_z^2 \langle \Delta^2 \rangle}.
    \label{eq:preliminfosig2}
\end{equation}
In this expression, the first term comes from the dependence of $\langle \Delta^2 \rangle$ (i.e., $\sigma^2_{\hat{\xi}_I}$) on $A_z$ (via $P_1$ and $\xi_I$) in Equation~\ref{eq:xiIvariance}; the second comes from the dependence of $\xi_I$ on $A_z$ via Equation~\ref{eq:xiIexpression}.

To calculate the first term, we must differentiate Equation~\ref{eq:xiIvariance}. Let $\langle\delta\rangle_B$ be the mean density in our bin $B$, which has width $\Delta\delta$. Then if $\Delta\delta$ is reasonably small, and writing $\delta$ for $\langle\delta\rangle_B$, we can approximate
\begin{equation}
    P_1 = \frac{1}{\sqrt{2\pi A_z \sigma^2_\mathrm{fid}}}\exp\left(\frac{-\delta^2}{2A_z\sigma^2_\mathrm{fid}}\right) \Delta \delta,
    \label{eq:P1}
\end{equation}
and thus
\begin{equation}
    \frac{d}{dA_z} P_1 = \frac{P_1}{2A_z} \left(\nu^2 - 1 \right).
    \label{eq:dP1}
\end{equation}

We already have $(d/dA_z) \xi_I = -\xi_I/A_z.$ Thus we derive, to first order in $\gamma$,
\begin{multline}
\frac{L^2}{2} = \frac{1}{8A_z^2(1-P_1)^2}\left\{ (P_1-2)^2(\nu^2-1)^2\vphantom{\frac12}\right.\\
    - \left.\frac{2\gamma\nu^2 (\nu^2 - 1)(P_1 - 2)\left( 4 P_1^2 + P_1(\nu^2-7) + 2\right)}{(1 - P_1)}\right\}
\end{multline}

Turning to the second term of Equation~\ref{eq:preliminfosig2}, we note that it is second-order in $\gamma$, and we therefore ignore it.\footnote{It is worth noticing however that this term is proportional to $N_p$, which will be extremely large. Thus its lack---and that of other second-order terms---will have a visible impact in Fig.~\ref{fig:infotest}.}
We thus obtain the following expression for the information on $A_z$ contained in one indicator function correlation:
\begin{multline}
    \mathcal{I}_{A_z} = \frac{1}{A_z^2(1-P_1)}\left\{ \frac{(P_1-2)^2(\nu^2-1)^2}{8(1-P_1)}\right.\\
    - \left.\frac{\gamma \nu^2 (\nu^2 - 1)(P_1 - 2)\left( 4 P_1^2 + P_1(\nu^2-7) + 2\right)}{4(1 - P_1)^2}\right\}.
\label{eq:biginfoeq}
\end{multline}

In practice, the second term of this expression is almost negligible, so we can write
\begin{equation}
    \mathcal{I}_{A_z} = \frac{1}{A_z^2(1-P_1)}\frac{(P_1-2)^2(\nu^2-1)^2}{8(1-P_1)}
\label{eq:sminfoeq}
\end{equation}

This expression is, counterintuitively, \emph{not} proportional to the volume, and we defer discussion of this fact to Section~\ref{sec:disc}. Before presenting our tests of this result, it is thus worthwhile to note the following caveats: Equation~\ref{eq:sminfoeq} assumes (1) that a large number of survey cells fall into the bin (Equation~\ref{eq:N1big}); (2) that Equation~\ref{eq:condvar} is a good approximation for the conditional variance; and (3) that the bin width $\Delta\delta$ is reasonably small (see Equation~\ref{eq:P1}).

\subsection{The Low-Probability Limit}
Whereas Equation~\ref{eq:sminfoeq} assumes the indicator function is unity on a significant number of survey cells ($N_1 \gg 1$), we can also write an expression for the information in the opposite limit ($N_1 \ll 1$), as follows.

Let us assume that $\mathcal{P}(\hat{N}_1 > 1)$ is small enough to ignore, so that $\hat{N}_1 \in \{0, 1\}$. In such a case, we can treat $\hat{N}_1$ as a Poisson variable with expected value $N_c P_1$, so that $\mathcal{P}(\hat{N}_1 = 1)=N_c P_1$ and $\mathcal{P}(\hat{N}_1 = 0)=1-N_c P_1$. In this case there can be no pairs of cells in the density bin, so that $\hat{N}_{11} = 0$ with probability one.

There are thus two possibilities for the $\hat{\xi}_I \equiv \hat{P}_{11}/\hat{P}_1^2 - 1$. If $\hat{N}_1 = 1$, then $\hat{\xi}_I = -1$, which occurs with probability $N_c P_1$; otherwise then $\hat{\xi}_I$ is undefined, with probability $1 - N_c P_1$.

Since we assume the underlying density field is Gaussian, $(d/d\sigma^2) P_1 = (P_1/2\sigma^2) (\nu^2 - 1)$. Summing over the two possibilities for $\hat{\xi}_I$, we obtain for the information (to first order in ${N}_1$)
\begin{equation}
    \mathcal{I}_{A_z} = \frac{N_1 (\nu^2-1)^2}{4 A_z^2}.
    \label{eq:twovalapprox}
\end{equation}

Since this approximation assumes $P_1 N_c \ll 1$, its validity is limited to
\begin{equation}
    \frac{\nu^2}{2} \gg \ln \Delta\nu N_c - \ln \sqrt{2\pi} \approx \ln \Delta\nu N_c - 1.
    \label{eq:twovaluevalidity}
\end{equation}

\section{Testing the result}
\label{sec:infotest}
We can now test Equations~\ref{eq:sminfoeq} and \ref{eq:twovalapprox} in two ways---first, by directly estimating the information from multiple realizations of a Gaussian field; second, by using measurements of an indicator correlation function to constrain the power spectrum amplitude.

\subsection{Measurements from Simulations}
\label{sec:infotestmeas}
For our first set of tests of Equation~\ref{eq:biginfoeq}, we measure the PDF of $\hat{\xi}_I$ from multiple realizations of a random Gaussian field, generated\footnote{Using the package \texttt{FyeldGenerator}, \citet{FyeldGen}.} from scalings of the Millennium Simulation linear power spectrum.

We first generate such random fields at $32^3$ grid points in a cubical volume (with periodic boundary conditions) of side length $500 h^{-1}$Mpc, simulating a survey with resolution of $\sim \!16h^{-1}$Mpc. Since our goal is to measure the effect of amplitude on $\mathcal{P}(\hat{\xi}_I)$, we use the variance as a proxy for amplitude (i.e., setting $\sigma^2_\textrm{fid} = 1$ in Equation~\ref{eq:Az}), performing 10,000 realizations with $\sigma^2 = 0.60$ and another 10,000 with $\sigma^2 = 0.65$. (These values approximate $\sigma_8 = 0.8$.) From these we can numerically estimate $(d/d\sigma^2)\mathcal{P}(\hat{\xi}_I)$ for various density bins (ranging from $\delta=-5.5$ to $5.5$, with bin width $\Delta\delta = 0.5$)\footnote{Since we assume $\delta$ is a Gaussian random variable, its domain is unbounded. In cosmological applications $\delta$ is non-Gaussian, and $\delta < -1$ is unphysical. In such cases we could perform the analysis with $A=\ln(1+
\delta)$, which has no lower bound.} and thus the information in $\xi_I$ for each of these bins. To estimate error bars for our information calculations, we perform 50 sets of such realizations (each set consisting of 10,000 realizations at each of the two values of $\sigma^2$).

In the same way, we also generate random fields at $64^3$ grid points in a cubical volume of side length $1000 h^{-1}$Mpc, (using the same values for $\sigma^2$), thus simulating a survey of eight times the volume.\footnote{The $32^3$-cell realizations are generated independently of the $64^3$-cell realizations, rather than being subsamples of the same.}

\begin{figure*}
\centering
\includegraphics[width=0.49\linewidth]{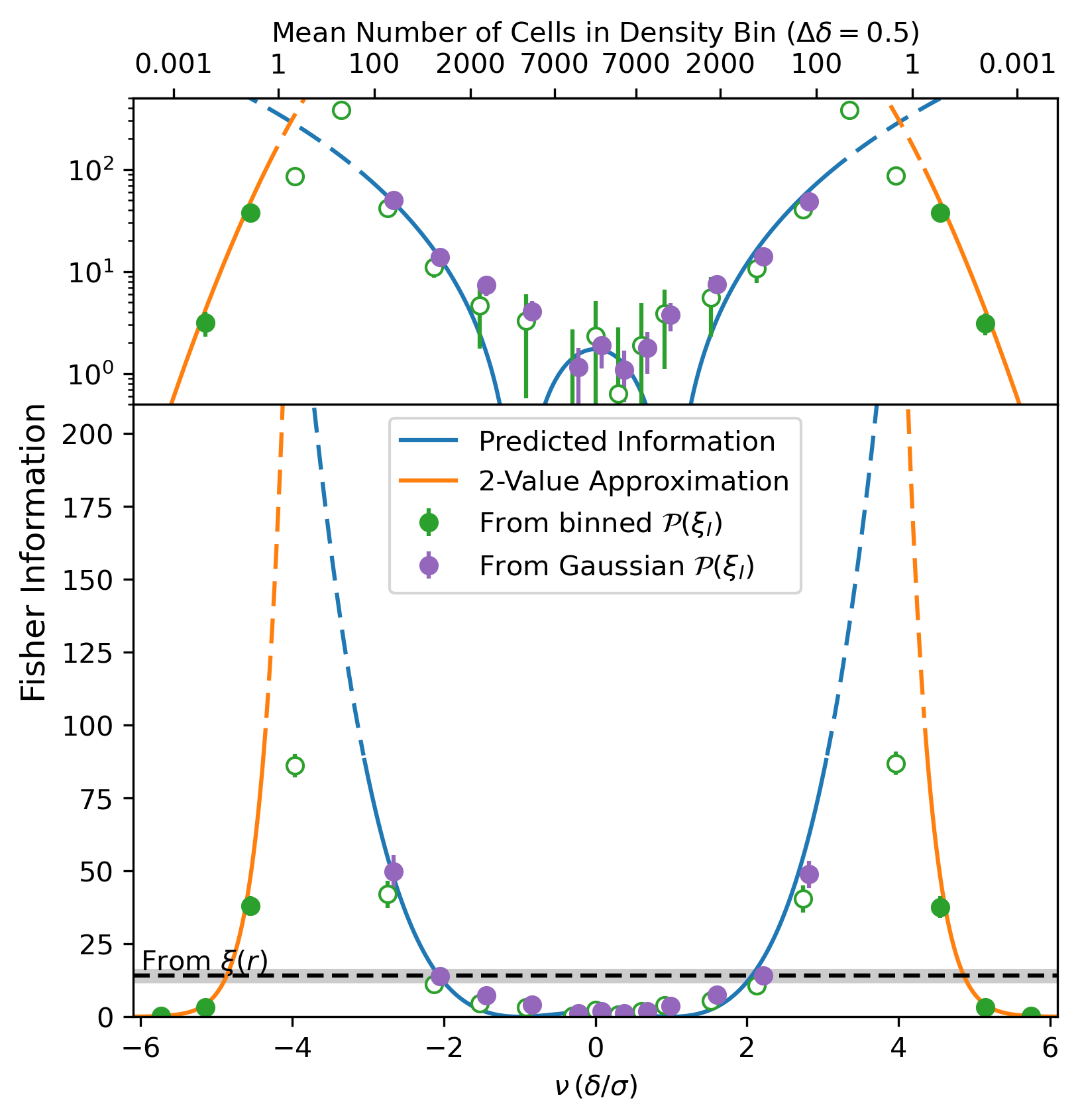}
\includegraphics[width=0.49\linewidth]{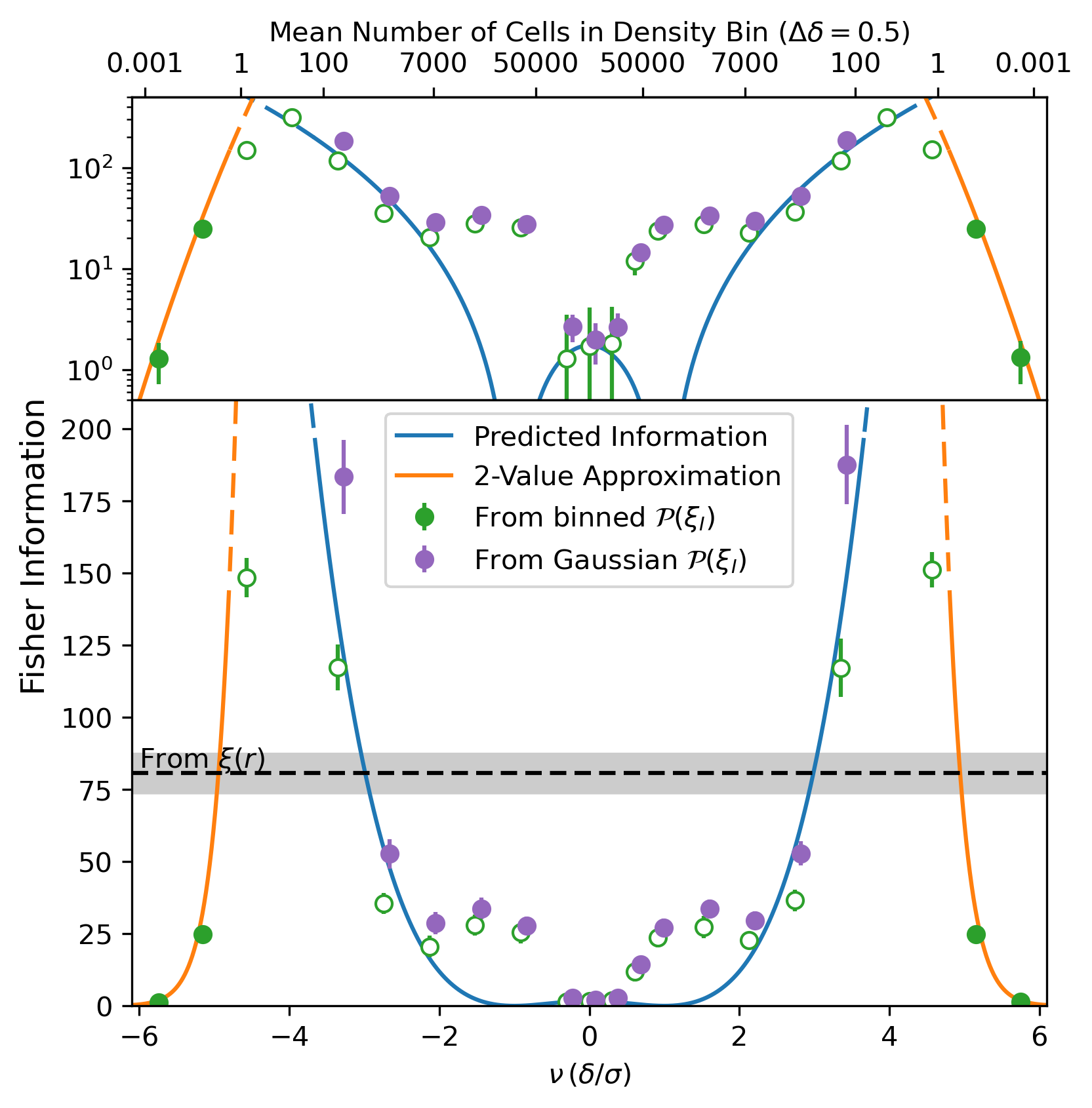}
\caption{Information on power spectrum amplitude (linear and log scales) from indicator correlation functions $\xi_I(r)$ in the distance bin $r \in [60, 80) h^{-1}$Mpc, from Gaussian realizations, for various densities $\nu = \delta/\sigma$. The left-hand panel shows results from a cube of 500$h^{-1}$ Mpc per side, divided into $32^3$-cells; the right-hand panel, from a cube of 1000$h^{-1}$Mpc per side, divided into $64^3$ cells. Both panels compare calculated information to the predictions of Equations~\ref{eq:biginfoeq} (blue) and \ref{eq:twovalapprox} (orange). Dashed portions of the curves indicate values of $\nu$ outside the formulas' applicability range. We derive the purple points by numerically differentiating a Gaussian approximation to the observed probability distribution of $\hat{\xi}_I$ (i.e., a continuous normal pdf with mean and variance given by $\langle \hat{\xi}_I\rangle$ and $\langle \hat{\xi}_I^2\rangle - \langle \hat{\xi}_I\rangle^2$). We derive the green points by binning the observed values of $\hat{\xi}_I$, taking the occurrence rate of each bin as the pdf, which we then numerically differentiate. See text (Section~\ref{sec:infotestmeas} and Appendix~\ref{sec:appendix}) for more information. The upper axis shows the expected number of survey cells in each bin. (Note that our use of 10,000 realizations allows us to resolve a mean number of cells down to $10^{-4}$.) Both panels also show the information from the full correlation function $\xi(r)$ for this particular distance bin.}
\label{fig:infotest}
\end{figure*}

In both cases, we determine $\hat{\xi}_I(r)$ where $r$ is in the radial distance bin $R = [60, 80)\,h^{-1}\textrm{Mpc}$, calculating it directly from $\hat{P}_{11}$ and $\hat{P}_1$ (Equation~\ref{eq:xihatdef}).

We determine the information on the amplitude in two ways. Both ways involve numerical estimation of $d\mathcal{P}/d\sigma^2$ as $(\mathcal{P}_2(\hat{\xi}_I) -\mathcal{P}_1(\hat{\xi}_I))/\Delta\sigma^2$, where the two values of $\mathcal{P}(\hat{\xi}_I)$ reflect the two sets of realizations (one set at each of the two given values of $\sigma^2$, with $\Delta\sigma^2 = 0.05$). The difference comes in how they determine the values of $\mathcal{P}(\hat{\xi}_I)$. In the regime for which Equation~\ref{eq:xiIdistrib} holds (i.e., where Equation~\ref{eq:sminfoeq} is valid), $\mathcal{P}(\hat{\xi}_I)$ is very nearly Gaussian, and so we can get well-behaved approximations for $\mathcal{P}(\hat{\xi}_I)$ by empirically measuring the mean and variance of $\hat{\xi}_I(r)$, from which we construct a Gaussian pdf. The Fisher-information values calculated from these Gaussian $\mathcal{P}(\hat{\xi}_I)$s appear as purple points in Fig.~\ref{fig:infotest}.\footnote{We also correct for the fact that the Gaussian is truncated at $\hat{\xi}_I = -1$. This correction is typically minimal, due to the limits explained in the next sentence.} We limit the use of this method to $\delta$-values for which (a) less than ten percent of the normal distribution (i.e., the Gaussian approximation to $\mathcal{P}(\hat{\xi}_I)$) would fall below $-1$, and (b) the expected number of cells $N_1$ is greater than 10 (pursuant to Equation~\ref{eq:N1big}). For comparison, the blue curves show the predictions of Equation~\ref{eq:biginfoeq} (solid where, based on the conditions in the previous sentence, the Gaussian approximation for $\mathcal{P}(\hat{\xi}_I)$ is valid, and dashed otherwise).

To test the low-probability expression of Equation~\ref{eq:twovalapprox}, we need another method for estimating $\mathcal{P}(\hat{\xi}_I)$. We perform this estimate by binning the measured $\hat{\xi}_I(r)$-values to obtain discrete $\mathcal{P}(\hat{\xi}_I)$s, without making any assumptions about the shape of the distribution. (See Appendix~\ref{sec:appendix} for our binning algorithm, as well as our correction for random fluctuations between sets of realizations.) The information values calculated from binning appear as green points in Fig.~\ref{fig:infotest}. For comparison, the orange curves show the predictions of Equation~\ref{eq:twovalapprox} (solid where $\nu^2/2 \ge \ln \Delta\nu N_c - 0.5$, see Equation~\ref{eq:twovaluevalidity}, dashed otherwise). Both figures use empty circles to denote results falling between the limits of the two predictions (i.e., where both curves are dashed).

Inspection of the figures shows that the approximations (such as working only to first order) made in deriving Equations~\ref{eq:sminfoeq} and \ref{eq:twovalapprox} do have an impact on the their accuracy; in particular it is evident that higher-order terms prevent information from vanishing at $\nu=\pm 1$, as Equation~\ref{eq:sminfoeq} would indicate. Nevertheless, it is clear that the greatest amount of information comes from $\nu$-values at or near the limit of the Gaussian approximation; in both tested cases, this corresponds to $N_1$ (the expected number of survey cells falling into the density bin---shown on the upper axis\footnote{Note that our use of 10,000 realizations allows us to resolve expected numbers of cells down to $10^{-4}$.}) being of order of magnitude of 100. At such densities, the cells are rare enough to be sensitive to changes in amplitude, but not so rare as to be dominated by Poisson noise.

The figure also shows the information contained in the standard correlation function $\xi(r)$, in the same radial distance bin $R=[60, 80)h^{-1}$ Mpc; we calculate this information by modeling the observed distribution of $\hat{\xi}(r)$ as a Gaussian. Two facts are evident: (a) near the limits of the Gaussian approximation (again, where $N_1$ is of order of magnitude $100$), the indicator correlation function $\xi_I(r)$ contains \emph{more} information than the full correlation function $\xi(r)$ at this distance; (b) the information in $\xi(r)$ does increase with volume (as expected),\footnote{Although it seems not to be completely proportional to volume; comparing the left panel to the right, the volume increases by a factor of 8, whereas the information increases only by 6; recall that in these tests we consider only one bin in $r$, not the information in the entire correlation function.}
 whereas, to first order, the information in $\xi_I(r)$ does not. Nevertheless, the larger volume allows Equation~\ref{eq:sminfoeq} to be valid for larger values of $|\nu|$; thus the information in $\xi_I(r)$ for these liminal $\nu$-values continues to outstrip the information in $\xi(r)$. We discuss both of these counter-intuitive observations below in Section~\ref{sec:disc}.

\subsection{Amplitude-Constraining Ability}

A second way to test our results is to see if indicator function correlations indeed allow us to constrain the amplitude to (at most) the inverse square root of the information. We can do the same with the full correlation function $\xi(r)$ in order to compare the information in each statistic at this distance. We perform this test with the $64^3$-cell realizations from the previous section.

\begin{figure*}
\centering
\includegraphics[width=0.49\linewidth]{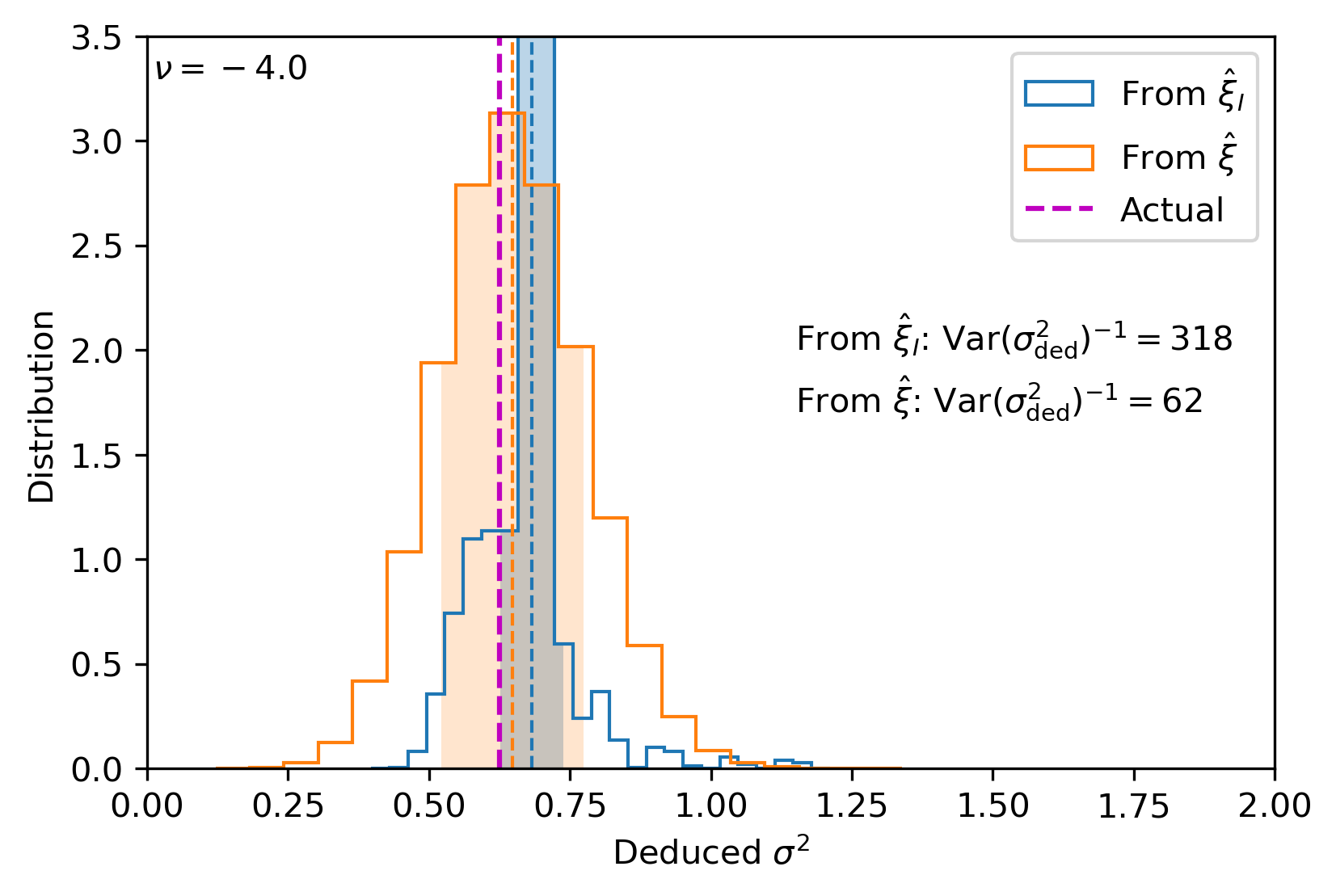}
\includegraphics[width=0.49\linewidth]{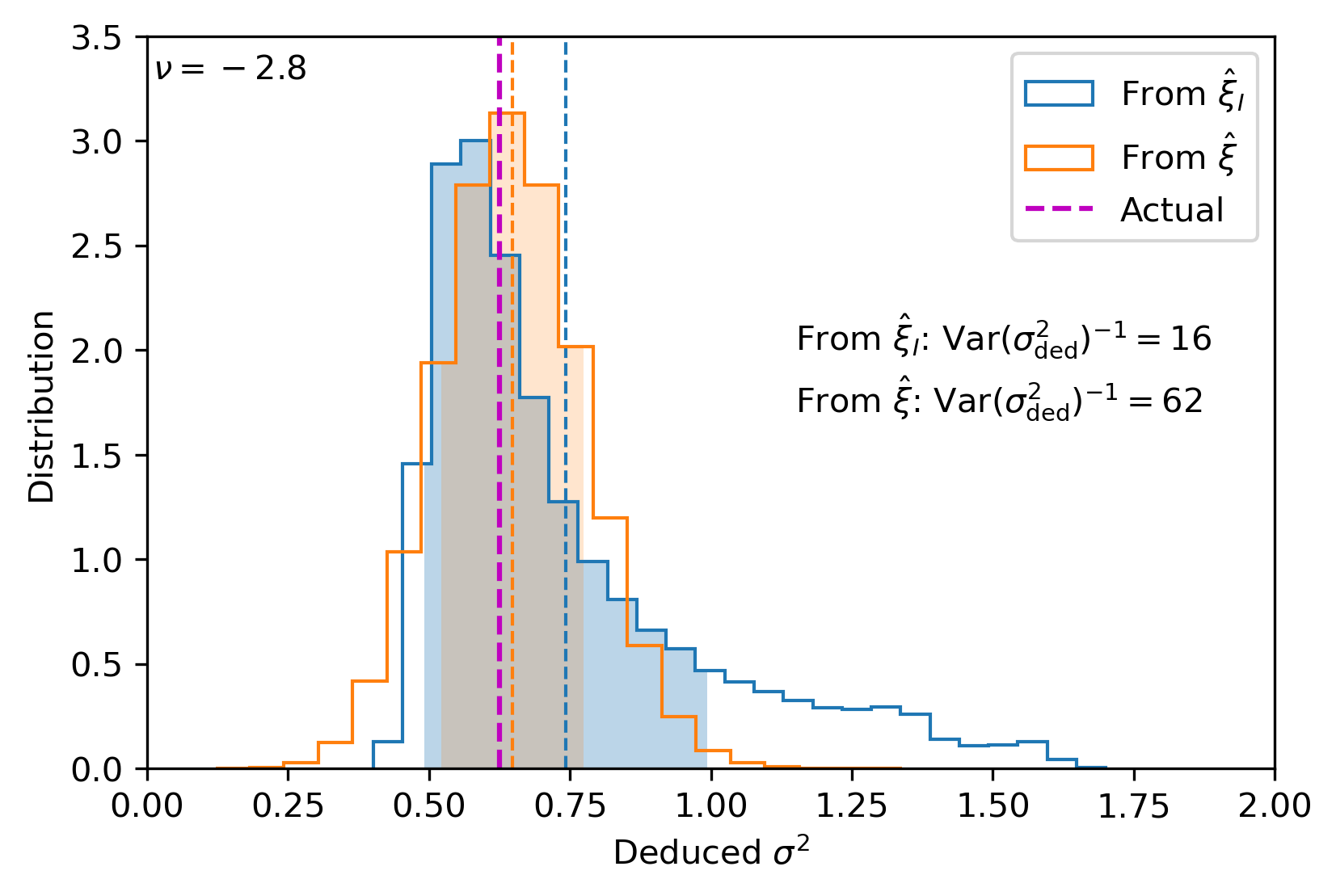}
\caption{Constraints on the value of $\sigma^2$ (proxy for amplitude $A_z$) deduced from indicator function correlations ($\xi_I$, blue) and the full correlation function ($\xi$, orange), in the set of $64^3$-cell realizations described above, in the radial distance bin $[60, 80)h^{-1}$ Mpc. Dashed blue and orange lines show mean deduced values, and the shading shows one standard deviation on either side of the mean; the magenta dashed line shows the actual value of ($\sigma^2=0.625$) employed to generate the realizations. Left panel: $\nu = \delta/\sigma \approx -4.0$; right panel: $\nu = \delta/\sigma \approx -2.8$. (Since our field is Gaussian, the sign of $\nu$ does not matter; cf. Equation~\ref{eq:sminfoeq}, which depends on $\nu^2$ only.)}
\label{fig:constraints}
\end{figure*}

Constraining the amplitude from the measured correlation function is straightforward: since we are assuming that we know $\gamma$, we obtain as the amplitude  $\sigma_\mathrm{deduced}^2 = \hat{\xi}(r)/\gamma$. Each realization thus yields its own deduced amplitude, and the orange histograms in Fig.~\ref{fig:constraints} show the distribution of these values (with their mean and the 1-$\sigma$ confidence interval).

Constraining the amplitude from the measured indicator correlation function $\hat{\xi}_I(r)$ requires more work. In Equation~\ref{eq:xiIexpression}, we could substitute the measured value of $\hat{\xi}_I$ for its expected value and write $\sigma^2_\mathrm{deduced} = \gamma \delta^2/\hat{\xi}_I$; however, since $\hat{\xi}_I$ scatters around its expected value, it is easy to find realizations in which (say) $\gamma$ is positive but $\hat{\xi}_I$ approaches zero (blowing the expression up) or, worse yet, is negative (yielding a negative $\sigma^2_\textrm{deduced}$). We shall thus use the measured value of $\hat{P}_1$ (the number of cells falling into the density bin) to impose a prior on $\sigma^2$. This does not introduce extra information, since we must measure $\hat{P}_1$ in order to calculate $\hat{\xi}_I$.

In particular, \citet{ReppSzapudi2021} demonstrate that the variance for a counts-in-cells probability is
\begin{equation}
    \sigma_1^2 \equiv \textrm{Var}(\hat{P}_1) = \frac{P_1(1-P_1)}{N_c} + \frac{N_c-1}{N_c}\overline{\xi}_I^{\neq} P_1^2,
\end{equation}
where, again, $\overline{\xi}_I^{\neq}$ is the volume-averaged correlation function, excluding same-cell correlations. But by Equation~\ref{eq:xiIexpression},
\begin{equation}
    \overline{\xi}_I^{\neq} = \frac{\delta^2}{\sigma^4}\overline{\xi}^{\neq},
\end{equation}
and since $\overline{\xi}^{\neq} < \sigma^2$, we conclude that
\begin{equation}
    \sigma_1^2 < \frac{P_1(1-P_1)}{N_c} + \frac{\delta^2}{\sigma^2} P_1^2.
\end{equation}

Since, given a $\delta$-bin, the variance completely determines the theoretical $P_1$, we can invert the relationship to write the inferred variance $\hat{\sigma}^2$ as a function of the observed counts-in-cells probability $\hat{P}_1$; thus,
\begin{align}
    \textrm{Var}(\hat{\sigma}^2) & = \textrm{Var}(\hat{P}_1) \left( \frac{\partial P_1}{\partial \sigma^2}\right)^{-2}\\
    & < \left( \frac{\hat{P}_1(1-\hat{P}_1)}{N_c} + \frac{\delta^2}{\sigma^2} \hat{P}_1^2 \right) \left(\frac{\hat{P}_1}{2\hat{\sigma}^2} \frac{\delta^2 - \hat{\sigma}^2}{\hat{\sigma}^2}\right)^{-2}.
    \label{eq:varsig2}
\end{align}

Hence, given the observed fraction of counts $\hat{P}_1$, we can infer a value for $\hat{\sigma}^2$ and then center the prior distribution (of the actual, underlying amplitude) upon this inferred value:
\begin{equation}
\mathcal{P}(\sigma^2) = \frac{1}{\sqrt{2\pi\textrm{Var}(\hat{\sigma}^2)}}\exp\left(-\frac{(\hat{\sigma}^2-\sigma^2)^2}{2\textrm{Var}(\hat{\sigma}^2)}\right),
\end{equation}
where we assume equality in Equation~\ref{eq:varsig2} as an estimate of $\textrm{Var}(\hat{\sigma}^2)$.

Furthermore, Equation~\ref{eq:xiIvariance} gives us the expected variance of the measured $\hat{\xi}_I$, allowing us to write
\begin{equation}
    \mathcal{P}(\xi_I | \sigma^2) = \frac{1}{2\pi\textrm{Var}(\hat{\xi}_I)}\exp\left( -\frac{(\xi_I - \hat{\xi}_I)^2}{2\textrm{Var}(\hat{\xi}_I)}\right).
\end{equation}

Hence, given a measured value for $\hat{\xi}_I$, we can write
\begin{equation}
    \mathcal{P}(\sigma^2|\hat{\xi}_I) \propto \mathcal{P}(\xi_I|\sigma^2) \mathcal{P}(\sigma^2),
    \label{eq:posterior}
\end{equation}
from which we can obtain the posterior $\sigma^2_\textrm{deduced}$, as the maximum likelihood value of the amplitude.

Summarizing, in each realization we can measure $\hat{P}_1$ in the course of measuring $\hat{\xi}_I$. From these measurements, Equations~\ref{eq:varsig2}--\ref{eq:posterior} let us deduce a posterior value for $\sigma^2_\mathrm{deduced}$  from $\hat{\xi}_I$, and the blue histograms in Fig.~\ref{fig:constraints} display these values (along with their mean and 1-$\sigma$ confidence interval).

When we compare Fig.~\ref{fig:constraints} with the right-hand panel of Fig.~\ref{fig:infotest}, we see that our method of constraining $\sigma^2$ is not optimally efficient (i.e., it does not necessarily achieve the Cram\'er-Rao bound). We expect this to be the case for $\xi_I$ because of the approximations we made (e.g., treating Equation~\ref{eq:varsig2} as an equality), but it is also the case for $\xi$: in this case the information is 80, whereas the inverse variance is 62. Nevertheless, this exercise demonstrates that in practice, use of the proper indicator function correlations can deliver tighter constraints on the amplitude than can the correlation function itself.

\section{Discussion} 
\label{sec:disc}
We previously noted two counter-intuitive features of Equation~\ref{eq:sminfoeq}. The first is that, for sufficiently large $|\nu|$-values, the indicator functions correlations can contain \emph{more} information than the correlation function within a fixed distance range. 
The left-hand panel of Fig.~\ref{fig:constraints} verifies this prediction, showing that indicator function analysis provided tighter constraints on the amplitude than analysis with the correlation function. This is in part due to the fact that indicator function correlations add counts-in-cells data to the clustering data, so one could expect it to provide more information. However, we must make sense of this feature in light of the fact that two-point correlations indeed capture all the information present in a zero-mean Gaussian field.

It is helpful to remember that in this case we are looking specifically at a particular distance bin $R$; to derive the full information we would need to consider all distance bins $R$ (and cross-correlations between them). This is complicated by the fact that different distance-bins, unlike $k$-modes, are not independent. Thus, we are not comparing \emph{all} the information inherent in these functions, but only the information at a given separation. What our result shows is, first, that the highest peaks (and lowest valleys) contain the most information, and, second, that $\xi$ averages the different densities in a non-optimal manner, diluting the analysis with non-informative survey cells. The indicator function correlations distill the information by focusing on the most-informative survey cells (see, e.g., fig.~4 and accompanying discussion in \citealp{ReppSzapudi2022}).

The other counter-intuitive feature is the lack of a volume term in Equation~\ref{eq:sminfoeq}; comparison of the two panels of Fig.~\ref{fig:infotest} shows that the placement of the blue curve is unchanged under an eight-fold volume increase. To some degree, this is the result of working to first order; the second term of Equation~\ref{eq:preliminfosig2} depends on $N_p$ (the number of pairs in the survey volume) but was dropped due to its dependence on $\gamma^2$. Inspection of the figure makes it evident that volume \emph{does} affect the placement of the purple points near $|\nu| = 1$.

However, a much more important reason for the volume-independence is our assumption in Equation~\ref{eq:N1big}, which limits the applicability of Equation~\ref{eq:sminfoeq}. Even though the placement of the blue curves in Fig.~\ref{fig:infotest} is unaffected by volume, that is not true for the regime of applicability (indicated by the transition from solid to dashed). This dependency reflects the paucity of high peaks in a volume-limited survey; the smaller the survey, the more rare---and thus the less informative---are the high peaks. On the other hand, larger survey volumes allow us to apply Equation~\ref{eq:sminfoeq} to higher values of $|\nu|$. Since $N_1$ is proportional to survey volume $V$, the cutoff value of $\nu^2$ depends on $\ln V$. Practically speaking, therefore, the information does depend on volume (albeit indirectly), since volume dictates the applicability of Equation~\ref{eq:sminfoeq}.

We can tie these two counter-intuitive features together as follows. By Equation~\ref{eq:sminfoeq}, the information scales roughly as $\nu^4$; but, when we fix a minimum value for $N_1$ (say, $N_1=100$), Equation~\ref{eq:N1big} produces a maximum value for $\nu^2$ depending on the logarithm of the survey volume. Therefore, the maximum information from $\xi_I(r)$ goes as $(\ln V)^2$. The information in $\xi(r)$, on the other hand, is roughly proportional to $V$ itself. Therefore, for any value of $\nu$, as the survey volume increases, the information in $\xi(r)$ will eventually outstrip the information available in any given $\xi_I(r)$. So in an infinite survey of a Gaussian field, no statistic could outperform $\xi(r)$; for actual surveys, however, $\xi_I(r)$ can often do better.

Our analysis has of course completely ignored the fact that, for realistic matter distributions, the highest peaks are non-linear. One can partially alleviate this non-linearity by performing the analysis on the log density \citep{Neyrinck2009,CarronSzapudi2013,ReppSzapudi2017}.

In addition, \citet{ReppSzapudi2022} suggest that indicator functions can selectively excise these nonlinearities, extending linear methods to higher $k$-values without the complications inherent in perturbation theory. Clipping \citep{Simpson2011,Simpson2013,Simpson2016} is a conceptually similar method of removing the non-linear peaks (see, e.g., \citealp{Lombriser2015,Giblin2018}). Recall that if $I_{p_i}(\mathbf{r})$ is the indicator function field determined by a specific density $p_i$, then the conventional overdensity  $\delta(\mathbf{r})$ is the weighted sum of these indicator functions,
\begin{equation}
    \delta(\mathbf{r})=\sum p_i\,I_{p_i}(\mathbf{r}),
\end{equation}
whereas the clipped field is
\begin{equation}
    \delta_C(\mathbf{r})=\sum_{p_i \le C} p_i\,I_{p_i}(\mathbf{r}).
\end{equation}
The results in this work suggest that it might be possible to extract the majority of the information from the highest-density bins ($p_i \approx C$) alone.

Finally, we note the relevance of this analysis to baryon acoustic oscillation (BAO) scales, on which the matter distribution approaches  Gaussianity. Our work thus has a straightforward application to measurement of the BAO amplitude. In addition, both \citet{Neyrinck2018} and \citet{Xu2025} show (using sliced correlation functions and density-split statistics, respectively---both closely related to indicator functions) that different density levels yield somewhat different positions for the BAO peak. Since these density-dependent shifts smear out the BAO feature, it follows that analysis at the optimal density levels could provide more accurate measurements of the BAO position, thus sidestepping the model-dependence inherent in most reconstruction methods.

\section{Conclusion}
\label{sec:concl}
In summary, when the parameter of interest is the amplitude of the power spectrum, it is straightforward to calculate the total Fisher information contained in an $n$-point Gaussian field.

However, this information is not evenly distributed among the $n$ cells; rather, certain densities are more information-rich than others. Using indicator functions to select the cells in a given density bin, we can calculate the amount of information (on the power-spectrum amplitude) in the indicator correlation function $\xi_I$ for a given separation distance $r$. Upon doing so, we find that we obtain the most information by considering cells with high absolute density contrast, as long as the survey contains enough such cells ($N_1$ of order of magnitude $100$, for the $60$--$80h^{-1}$-Mpc distance bin we considered). For typical surveys one can even find density bins for which the indicator correlation function $\xi_I(r)$ contains more information than the full correlation function $\xi(r)$ (at a specified separation distance $r$), simply because indicator function analysis weights the cells in a more-nearly optimal manner.

Although this work focuses on Gaussian fields (and has not attempted to address discrete sampling), future work will seek to extend these results to more realistic distributions. We conclude that indicator function analysis is a promising strategy for extracting maximal information from cosmological survey data. 

\section*{Acknowledgments}
IS acknowledges NASA grant 
23-ADAP23-0061. 
YC acknowledges the support of the UK Royal Society through a University Research Fellowship. For the purpose of open access, the author has applied a Creative Commons Attribution (CC BY) license to any Author Accepted Manuscript version arising from this submission. This research was supported in part by grant NSF PHY-2309135 to the Kavli Institute for Theoretical Physics (KITP). The authors also acknowledge the valuable comments received from the anonymous reviewer.

\section*{Data availability}
The scripts employed in generation and analysis of the Gaussian simulations used in this paper, as well as the data recorded from each simulation, are available upon reasonable request to the authors.

\bibliographystyle{astron}
\bibliography{indicators_mnras}

\appendix
\section{Binning Details}
\label{sec:appendix}
In binning the values of $\hat{\xi}_I$ (for determining the green points in Fig.~\ref{fig:infotest}), there are cases in which we want every $\hat{\xi}_I$-value to have its own bin (the most extreme of these cases being the extreme densities for which $\hat{\xi}_I$ is either undefined or $-1$ in any given realization). At the same time, if we have too many bins, shot noise will mimic true $\partial \mathcal{P}(\xi_I)/\partial A_z$.

Thus we use the following binning algorithm: suppose we have $N_r$ realizations. Since $\hat{\xi}_I$ can be undefined, we will have $N_r^\textrm{def}$ instances of defined $\hat{\xi}_I$-values and $N_r - N_r^\textrm{def}$ cases in which it is undefined. Let $n$ be the number of unique numerical (defined) values of $\hat{\xi}_I$ in this set. We begin with the \emph{lower} of the two values $\sqrt{N_r^\textrm{def}}$ and $n$, and split the range of $\hat{\xi}_I$ into that many bins; this is our minimum number of bins. If all of these bins are full, we stop; if some of these bins are empty, we double the number of bins until either we have $\sqrt{N_r^\textrm{def}}$ full bins, or each unique value occupies its own bin. We treat cases in which $\hat{\xi}_I$ is undefined as occupying their own non-numerical bin, as when deriving Equation~\ref{eq:twovalapprox}.

This procedure effectively errs on the side of having too many bins, thus leaving us open to shot noise masquerading as information. We attempt to correct for this ``pseudo-information'' as follows: we rerun the two sets of 10,000 realizations, but we use the same value of $\sigma^2=0.625$ for each. Any differences between the resulting $\mathcal{P}(\hat{\xi}_I)$ distributions will thus be due to random fluctuations, not to any modulation of $\sigma^2$. We then calculate the ``information'' as before,\footnote{I.e., by determining $\langle (d/d\sigma^2 \ln \mathcal{P}(\hat{\xi}_I))^2\rangle$, under the assumption of $\Delta \sigma^2 = 0.05$.} obtaining a numerical value for the shot-noise pseudo-information. We perform this procedure ten times to estimate the expected value and standard deviation of this pseudo-information. Finally, to obtain the green points in Fig.~\ref{fig:infotest}, we subtract off the expected pseudo-information value; to obtain the error bars, we add in quadrature the standard deviation to the existing uncertainty.

\bsp	
\label{lastpage}
\end{document}